\documentclass{article}

\newcommand{\forfinal}[1]{#1}

\forfinal{\usepackage{spconf}}
\usepackage{epsfig}

\usepackage{citesort}
\ifx\pdfoutput\undefined
\usepackage{graphicx}
\else
\fi

\usepackage{psfrag}
\usepackage{dsfont}
\usepackage{subfigure}

\usepackage{url}

\usepackage{stfloats}
\usepackage{amsmath,amsfonts,amssymb,amsxtra}
\usepackage{array}

\usepackage{pstricks,pst-node,pst-text,pst-3d,pst-plot,pst-all}

\newcommand{\E}[1]{{\rm E} {}}

\newcommand{\ist}{\hspace*{.2mm}}
\newcommand{\rmv}{\hspace*{-.2mm}}
\newcommand{\remark}[1]{}

\newcommand{\x}{{\bf x}}

\def\MSE{\varepsilon(\mathbf{x}; \hat{\mathbf{x}})}

\newtheorem{theorem}{Theorem}

\newcommand{\eq}{\,=\,}
\renewcommand{\jmath}{j}

\newcommand{\be}{\begin{equation}}
\newcommand{\ee}{\end{equation}}

\newcommand{\qed}{\nobreak \ifvmode \relax \else
      \ifdim\lastskip<1.5em \hskip-\lastskip
      \hskip1.5em plus0em minus0.5em \fi \nobreak
      \vrule height0.75em width0.5em depth0.25em\fi}

\DeclareMathOperator{\supp}{supp}

\forfinal{\parindent 1.0em}

\title{On \hspace*{-.7mm} Unbiased \hspace*{-.7mm} Estimation \hspace*{-.7mm} of Sparse Vectors corrupted by Gaussian noise}
\forfinal{\name{Alexander Jung$^a$\rmv, Zvika Ben-Haim$^b$\rmv, Franz Hlawatsch$^a$\rmv, and Yonina C.\ Eldar$^b$\thanks{\hspace*{-5mm}This work was supported by the FWF under Grant S10603-N13 (Statistical Inference) within the National Research Network SISE, by the WWTF under Grant MA 07-004 (SPORTS), by the Israel Science Foundation under Grant 1081/07, and by the European Commission under the FP7 Network of Excellence in Wireless COMmunications NEWCOM++ (contract no. 216715).}\vspace{-2.5mm}}
\address{\normalsize $^a$Institute of Communications and Radio-Frequency Engineering, Vienna University of Technology\\[-1.2mm]
\normalsize Gusshausstrasse 25/389, A-1040 Vienna, Austria; e-mail: \{ajung, fhlawats\}@nt.tuwien.ac.atÊ\\[1mm]
\normalsize $^b$Technion---Israel Institute of Technology, 
\normalsize Haifa 32000, Israel; e-mail: \{zvikabh@tx, yonina@ee\}.technion.ac.il}}

\begin{document}
\maketitle
\forfinal{\renewcommand{\baselinestretch}{0.935}\small\normalsize}

\begin{abstract}
We consider the estimation of a sparse parameter vector from measurements corrupted by
white Gaussian noise.
Our focus is on unbiased estimation as a setting under which the difficulty of the problem can be quantified analytically.
We show that there are infinitely many unbiased estimators but none of them has uniformly minimum mean-squared error.
We then provide lower and upper bounds on the Barankin bound, which describes the performance achievable by unbiased estimators.
These bounds are used to predict the threshold region of practical estimators.
\end{abstract}
\forfinal{}

\section{Introduction}\label{sec:intro}

\vspace{-1mm}

We consider the estimation of a parameter vector $\mathbf{x}_0 \!\in\! \mathbb{R}^{N}\rmv$
which is known to be $S$-sparse, i.e., at most $S \!\ge\!1$ of its entries are nonzero, where  typically $S \!\ll\! N$. This will be written as
$\mathbf{x}_0 \!\in\! \mathcal{X}_S \triangleq \{ \mathbf{x} \!\in\! \mathbb{R}^{N} \!\rmv : \rmv {\| \mathbf{x} \|}_{0} \!\leq\! S \}$,
where ${\| \mathbf{x} \|}_{0}$ denotes the number of nonzero elements of $\mathbf{x}$. Our focus is the frequentist estimation setting in which the parameter $\mathbf{x}_0$ is modeled as unknown but deterministic \cite{LC}.
While the sparsity $S$ is assumed known, the positions of the nonzero entries of $\mathbf{x}_0$ are unknown.
The $S$-sparse vector $\mathbf{x}_0$ is corrupted by white Gaussian noise with known variance $\sigma^{2}\rmv$, so that
the observation is
\begin{equation}
\label{equ_stat_model}
\mathbf{y} \ist=\ist \mathbf{x}_0 +  \mathbf{n} \,, \quad\; \text{with} \;\; \mathbf{x}_0 \rmv\in\rmv \mathcal{X}_S \ist , \;\mathbf{n} \rmv\sim\rmv \mathcal{N}( \mathbf{0}, \sigma^{2} \mathbf{I}_N ) \,.
\end{equation}
This will be called the \emph{sparse signal-in-noise model} (SSNM).

In this paper, we are interested in unbiased estimators, i.e., estimators $\hat{\mathbf{x}}(\mathbf{y})$ which satisfy
$\mbox{E}_{\mathbf{x}} \{ \hat{\mathbf{x}}(\mathbf{y}) \} = \mathbf{x}$ for
all $\mathbf{x} \!\in\! \mathcal{X}_S$.
(The notation $\mbox{E}_{\mathbf{x}}$ indicates that the expectation uses the distribution of $\mathbf{y}$ induced by the parameter value $\mathbf{x}$.)
The assumption of unbiasedness is commonly applied when analyzing estimation performance \cite{LC,ZvikaSSP}.
Apart from being intuitively appealing, unbiased estimation is known to be optimal when the signal-to-noise ratio (SNR) is high \cite{LC}.
More practically, to analyze achievable performance, the class of estimators must be restricted in some way in order
to exclude approaches such as $\hat{\mathbf{x}}(\mathbf{y}) \equiv \mathbf{x}_{1}$, which yield zero error at the
specific point $\mathbf{x} \!=\! \mathbf{x}_{1}$ but are evidently undesirable.
Consequently, even though many estimators for the SSNM are biased \cite{Mallat98},
we adopt the commonly used unbiasedness assumption in order to
gain insight into the achievable performance in this setting. Note that while $\mathbf{x}_0$ is known to be $S$-sparse,
the estimates $\hat{\mathbf{x}}$ we consider are not constrained to be $S$-sparse.

Our goal
is to characterize the minimum mean-squared error (MSE) achievable by unbiased estimators. This optimal MSE, known as the \emph{Barankin bound} (BB) \cite{Barankin}, cannot be calculated analytically. Instead, we prove that the BB is itself bounded above and below by simple, closed-form expressions. The upper bound is obtained by solving a constrained optimization problem, while the lower bound is based on the Hammersley--Chapman--Robbins technique \cite{HCRGormanHero}.
In a numerical study, we
demonstrate the ability of our bounds to roughly predict the threshold SNR of practical estimators.

A well-known lower bound on the BB is the Cram\'er--Rao bound (CRB), which was recently derived for the sparse setting \cite{ZvikaSSP,ZvikaCRB}.
To obtain the CRB, it is sufficient to assume that the estimator $\hat{\mathbf{x}}(\mathbf{y})$ is unbiased only at a given parameter value $\mathbf{x}_{0} \!\in\! \mathcal{X}_S$
and in its local neighborhood.
By contrast, in this paper we seek to characterize the MSE obtainable by estimators which are unbiased for \emph{all} values $\mathbf{x} \!\in\rmv \mathcal{X}_S$.
One may hope to achieve a higher---hence, tighter---lower bound
by adopting this more restrictive constraint.
Furthermore, the CRB is discontinuous in $\mathbf{x}$ \cite{ZvikaCRB}, whereas the MSE of an estimator for
the SSNM is always continuous \cite{LC}; this again suggests that it may be possible to find improved lower bounds.

We note in passing that our results can be easily generalized
to the sparse orthonormal signal model
\begin{equation}
\label{equ_stat_model_H}
\mathbf{y} \ist=\ist \mathbf{H} \mathbf{x}_0 + \mathbf{n} \,, \quad\; \text{with} \;\;
\mathbf{x}_0 \rmv\in\rmv \mathcal{X}_S \ist , \;\mathbf{n} \rmv\sim\rmv \mathcal{N}( \mathbf{0}, \sigma^{2} \mathbf{I}_M ) \,,
\end{equation}
where $\mathbf{y} \!\in\! \mathbb{R}^{M}\rmv$ with $M \rmv\ge\rmv N$ and the known matrix $\mathbf{H} \!\in\! \mathbb{R}^{M \times N}\rmv$
has orthonormal columns, i.e., $\mathbf{H}^{T} \mathbf{H} = \mathbf{I}_N$.
However, for simplicity of notation, we will continue to refer to the model \eqref{equ_stat_model}.
This setting arises, for example, in image denoising based on the assumption of sparsity in the wavelet domain.

\vspace{-3.5mm}

\section{Unbiased Estimation} \label{sec_unbiased_estimation}

\vspace{-.5mm}

We will denote the set of unbiased estimators by
\[
\mathcal{U} \,\triangleq\, \big\{ \hat{\mathbf{x}}(\cdot)  \, : \,
 \mbox{E}_{\mathbf{x}} \{ \hat{\mathbf{x}} ( \mathbf{y} )  \} \rmv=\rmv \mathbf{x}
 \ \text{ for all } \ \mathbf{x} \rmv\in\! \mathcal{X}_S
 \big\} \ist.
\]
Before analyzing bounds on the performance of unbiased estimators, we must ask whether such techniques exist at all. This is answered affirmatively in the following theorem. Due to space limitations, the proof of this and subsequent theorems is omitted and will appear in \cite{SparseToAppear}.
The MSE $\mbox{E}_{\mathbf{x}} \big\{ \|Ê\hat{\mathbf{x}}(\mathbf{y}) - \mathbf{x} \|^{2}_{2} \big\}$ of a given estimator $\hat{\mathbf{x}}(\cdot)$
at the parameter value $\mathbf{x}$ will be denoted $\MSE$.

\begin{theorem}
For $S\!=\!N$ in the SSNM \eqref{equ_stat_model}, there is only one unbiased estimator of $\mathbf{x}_0$ (up to
modifications having zero measure) which has a bounded MSE\@. This is the ordinary least-squares (LS) estimator,
\begin{equation}
\label{equ ls}
\hat{\mathbf{x}}_{\text{\em LS}}(\mathbf{y}) \,\triangleq\, \arg\min_{\mathbf{x} \in \mathbb{R}^{N}} \!\big\{ \|\mathbf{y} \!-\! \mathbf{x} \|_2^2 \big\} \rmv\eq \mathbf{y} \,,
\vspace{-.5mm}
\end{equation}
whose MSE is $\varepsilon(\mathbf{x}; \hat{\mathbf{x}}_{\text{\em LS}}) \equiv N\sigma^2$.
By contrast, for $S \!<\! N$, there are infinitely many unbiased estimators of $\mathbf{x}_0$.
\end{theorem}
The assumption of a bounded MSE means that there exists a constant $C$ such that $\MSE \leq C$ for all $\x$. This assumption is required to exclude certain pathological functions which do not make sense as estimators.
Since for $S\!=\!N$ the only unbiased estimator with bounded MSE is the LS estimator, we will hereafter assume that $S \!<\! N$.

\vspace{-1mm}

\section{Optimal Unbiased Estimation} \label{sec_optimum_unbiased_estimation}

\vspace{-.5mm}

Given a fixed parameter value $\mathbf{x}_0 \!\in\rmv \mathcal{X}_S$, one can define an optimal unbiased estimator
$\hat{\mathbf{x}}^{(\mathbf{x}_0)}(\cdot)$ for the SSNM \eqref{equ_stat_model} as an unbiased estimator which
minimizes the MSE at $\mathbf{x}_0$, i.e.,
\begin{equation}
\label{equ_opt_problem_1}
\varepsilon(\mathbf{x}_0; \hat{\mathbf{x}}^{(\mathbf{x}_0)}) \,= \min_{\hat{\mathbf{x}}(\cdot) \in\ist \mathcal{U}}
  \mbox{E}_{\mathbf{x}_0} \rmv\big\{ \|\hat{\mathbf{x}}(\mathbf{y}) - \mathbf{x}_0\|_2^{2} \big\} \,.
\end{equation}
Solving \eqref{equ_opt_problem_1}
is equivalent to solving the $N$ individual optimization problems
\begin{equation}
\label{equ_opt_problem_2}
\min_{\hat{x}_{k}(\cdot) \in\ist \mathcal{U}'} \rmv \mbox{E}_{\mathbf{x}_0} \rmv\big\{ |\hat{x}_{k}(\mathbf{y}) - x_{0,k}|^{2} \big\} \,, \quad k=1,...,N \,,
\end{equation}
where $\hat{x}_{k}(\cdot)$ and $x_{0,k}$ are the $k$th elements of $\hat{\mathbf{x}}(\cdot)$ and $\mathbf{x}_0$, respectively, and $\mathcal{U}'$ denotes the set of unbiased estimators of the $k$th element of $\mathbf{x}$.

Since the MSE $\MSE$
of an
estimator $\hat{\mathbf{x}}(\cdot)$ depends on the true parameter value $\mathbf{x} \!\in\! \mathcal{X}_S$, an unbiased
estimator that has a small MSE for one parameter value $\mathbf{x}_0 \!\rmv\in\! \mathcal{X}_S$ may have a large MSE
for other parameter values in $\mathcal{X}_S$. Sometimes, however, a single estimator
solves \eqref{equ_opt_problem_1} simultaneously for all $\mathbf{x} \!\in\rmv \mathcal{X}_S$. 
Such a technique
is called a \emph{uniformly minimum variance unbiased} (UMVU) estimator \cite{LC}. (Note that for an unbiased estimator the variance
equals the MSE, so minimum MSE implies minimum variance.)
However, for the SSNM \eqref{equ_stat_model}, the following negative result can be shown \cite{SparseToAppear}.

\begin{theorem}
For the SSNM with $S \!<\! N$, there does not exist a UMVU estimator.
\end{theorem}

Thus, no single unbiased estimator achieves minimum MSE for all values of $\x \!\in\! \mathcal{X}_S$.
We can still try to find a \emph{locally} minimum variance unbiased (LMVU) estimator by solving \eqref{equ_opt_problem_1} for a \emph{specific} $\mathbf{x}_{0} \!\in\rmv \mathcal{X}_S$.
This approach yields the unbiased estimator which has minimum MSE for the given parameter value $\x_0$. While the resulting technique may have poor performance for other parameter values, its MSE at $\x_0$ is of interest as it provides a lower bound
on the MSE at $\x_0$ of any unbiased estimator. This lower bound is known as the BB \cite{Barankin}. Because it is achieved, by some technique, at each point $\x_0$, the BB is the largest (thus, tightest) lower bound for unbiased estimators.
Unfortunately, an analytical expression of the BB is usually not obtainable.
Instead, in the following two subsections, we will provide lower and upper bounds on the BB.

\subsection{Lower Bound on the MSE}
\label{sec_lowerbound}

A variety of techniques exist for developing lower bounds on the MSE of unbiased estimators. The simplest of these is the CRB, which was derived for a more general
sparse estimation setting in \cite{ZvikaCRB,ZvikaSSP}. In our case, the CRB is given by
\begin{equation}
\label{equ crb}
\varepsilon(\mathbf{x}_{0};\hat{\mathbf{x}}) \ist\geq
\begin{cases}
S \sigma^2, & {\|\x_0\|}_0 = S \\[.8mm]
N \sigma^2, & {\|\x_0\|}_0 < S \ist .
\end{cases}
\end{equation}

The CRB assumes unbiasedness only in a neighborhood of $\x_0$.
Since we are interested in estimators which are
unbiased for all $\x \!\in\! \mathcal{X}_S$, we can expect to obtain higher bounds.
One such approach
is the Hammersley--Chapman--Robbins bound (HCRB) \cite{HCRGormanHero}, which can be formulated, in our context, as follows.
For a given parameter value $\mathbf{x}_{0}$, consider a set of $p$ ``test points''
$\{ \mathbf{v}_{i} \}_{i=1}^p$
such that $\mathbf{v}_i + \x_0 \in \mathcal{X}_S$.
The HCRB states that the MSE of any unbiased
estimator $\hat{\mathbf{x}}(\cdot) \!\in\rmv \mathcal{U}$ satisfies
\[
\varepsilon(\mathbf{x}_{0};\hat{\mathbf{x}}) \ist\geq\ist \mbox{Tr} \big( \mathbf{V} \mathbf{J}^{\dagger} \mathbf{V}^{T} \big) \,,
\]
where $\mathbf{V} \triangleq [ \mathbf{v}_{1} \cdots \mathbf{v}_{p} ] \!\in\! \mathbb{R}^{N \times p}$,
the elements of the matrix $\mathbf{J} \!\in\! \mathbb{R}^{p \times p}$ are given by
$(\mathbf{J})_{i,j} \triangleq \exp(\mathbf{v}^{T}_{i} \mathbf{v}_{j}/\sigma^{2}) - 1$,
and $\mathbf{J}^{\dagger}$ denotes the pseudoinverse of $\mathbf{J}$.

Both the test points $\mathbf{v}_i$ and their number $p$ are arbitrary and can depend on $\x_0$. Clearly, the challenge is to choose test points
which result in a tight but analytically tractable bound. To this end, we note the following facts. First, it can be shown \cite{SparseToAppear} that the CRB is obtained as a limit of HCRBs by choosing the test points $\mathbf{v}_{i}$ as\footnote{Hereafter, 
with a slight abuse of notation, the index $i$ of $\mathbf{v}_{i}$ is allowed to take on $p$ different values from the set
$\{ 1,\dots,N\}$.} 
\begin{subequations}\label{equ test pts crb}
\begin{align}
{\{ t \ist\mathbf{e}_i \}}_{i \ist\in\ist \supp(\x_0)} \ist, \quad \text{if }\ist{\|\x_0\|}_0 & \rmv=\rmv S
\label{equ test pts =s} \\[.5mm]
{\{ t \ist\mathbf{e}_i \}}_{i \ist\in\ist \{1,\ldots,N\}} \ist, \quad \text{if }\ist{\|\x_0\|}_0 &\rmv<\rmv S
\label{equ test pts <s}
\end{align}
\end{subequations}
and letting $t \rmv\rightarrow\rmv 0$.
Here, $\mathbf{e}_i$ represents the $i$th column of the $N \!\times\! N$ identity matrix and
$\supp(\x_0)$ denotes the set of indices of all nonzero elements of $\x_0$.
Second, the CRB \eqref{equ crb} is tight whenever ${\|\x_0\|}_0 \!<\rmv S$,
since in this case the bound is achieved by the LS estimator \eqref{equ ls}.
The CRB is also tight when ${\|\x_0\|}_0 \rmv= S$ and all components of $\x_0$ are
much larger (in magnitude) than $\sigma$ \cite{ZvikaCRB}.
Any laxity in the CRB will therefore arise when $\x_0$ contains exactly $S$ nonzero components, one or more of which are small.
For such $\x_0$, it follows from \eqref{equ test pts =s} that only $S$ test points are employed;
yet when the
\pagebreak 
small components of $\x_0$ are replaced with zero, then according to \eqref{equ test pts <s},
a much larger number $N$ of test points is used. In light of this, one could attempt to improve the CRB by increasing the number of test points when ${\|\x_0\|}_0 = S$. This should be done such that when the smallest component tends to zero, the test points converge to \eqref{equ test pts <s}, for which the CRB is tight. In doing so, one still needs to ensure that $\mathbf{v}_i + \x_0 \in \mathcal{X}_S$.

A reasonable choice which satisfies these requirements is
\begin{equation}
\label{equ_test_points}
\mathbf{v}_i =
\begin{cases}
t \ist \mathbf{e}_i \ist, & i \in \supp(\x_0) \\[.5mm]
-\ist x_{0}^{(S)} \mathbf{e}_k + t \ist\mathbf{e}_i \ist, & i \notin \supp(\x_0)
\end{cases}
\vspace{-.5mm}
\end{equation}
for $i=1,\ldots,N$. Here, $x_{0}^{(S)}$ is the $S$-largest (in magnitude) entry of $\x_{0}$ (note that $x_{0}^{(S)} \!=\rmv 0$ if $\ist{\|\x_0\|}_0 \rmv<\rmv S$). The test points \eqref{equ test pts crb}, which yield the CRB, are a subset of the choice of test points in \eqref{equ_test_points}. Consequently, it can be shown that the resulting bound will always be at least as tight as the CRB\@.

In analogy to the CRB, a simple lower bound can be obtained from \eqref{equ_test_points} by letting $t \rightarrow 0$.
Some rather tedious calculations \cite{SparseToAppear} yield the following result.

\begin{theorem}
\label{thm_lower}
The MSE of any unbiased estimator $\hat{\mathbf{x}}(\cdot) \!\in\! \mathcal{U}$ is bounded below
\vspace{-1mm}
by
\begin{equation}
\label{equ_HCR}
\varepsilon(\mathbf{x}_{0};\hat{\mathbf{x}}) \ist\geq
\begin{cases}
S\sigma^2 + (N \!-\! S \!-\! 1) \ist \sigma^2 \ist e^{-\xi_{0}^{2}/\sigma^{2}} \!, & {\|\x_0\|}_0 \rmv=\rmv S \\[.8mm]
N\sigma^2 \rmv, & {\|\x_0\|}_0 \rmv<\rmv S \,,
\end{cases}
\vspace{-1.5mm}
\end{equation}
where $\xi_{0} \triangleq \min_{k \in \supp(\mathbf{x}_{0})} |x_{0,k}|\ist$.
\end{theorem}

Strictly speaking, this bound
is obtained as a limit of HCRBs. However, for simplicity we will continue to
refer to \eqref{equ_HCR} as an HCRB\@.
The bound is higher (tighter) than the CRB \eqref{equ crb}, except when the CRB itself is already tight.
Furthermore, while both bounds are discontinuous in the transition between ${\|\x_0\|}_0 = S$ and ${\|\x_0\|}_0 \rmv<\rmv S$,
the jump in \eqref{equ_HCR} is much smaller than that in \eqref{equ crb}. This discontinuity can be eliminated altogether by adding more test points, but the resulting bound no longer has a simple closed form and can only be evaluated numerically.

\vspace{-1mm}
\subsection{Upper Bound on LMVU Performance}
\label{sec_upperbound}

We next derive an upper bound on the BB by approximately solving \eqref{equ_opt_problem_2} for $k = 1,\ldots,N$. When $k \in \supp(\mathbf{x}_0)$, it can be shown \cite{SparseToAppear} that the solution of \eqref{equ_opt_problem_2} is given by the
$k$th component of the LS estimator in \eqref{equ ls}, i.e.,
\begin{equation} \label{equ BBc in supp}
\hat{x}_k(\mathbf{y}) = y_k.
\end{equation}
It remains to consider $k \notin \supp(\mathbf{x}_{0})$.
For such $k$, let us denote $\hat{x}_k(\mathbf{y}) = y_{k} + \hat{x}_k^\prime (\mathbf{y})$.
To obtain an upper bound on the solution of the optimization problem \eqref{equ_opt_problem_2}, we will restrict the set of allowed estimators
$\hat{x}_k(\cdot)$ by requiring the correction component $\hat{x}_k^\prime(\cdot)$ to satisfy the following two properties.

\begin{itemize}
\item {\em Odd symmetry} with respect to index $k$ and all indices in $\supp(\mathbf{x}_{0})$:
For all $l \in \{k\} \cup \supp(\mathbf{x}_{0})$,
\begin{equation}
\hat{x}_k^\prime (\ldots,- y_{l}, \ldots) \ist=\ist -\ist \hat{x}_k^\prime ( \ldots, y_{l}, \ldots) \,.
\label{equ_constr_odd_symm}
\end{equation}
\item
{\em Independence} with respect to all other indices:
For all $\, l \notin \{k\} \cup \supp(\mathbf{x}_{0})$,
\vspace*{-.5mm}
\begin{equation}
\hat{x}_k^\prime (\ldots, y_{l}, \ldots) \ist=\ist \hat{x}_k^\prime ( \ldots, 0, \ldots) \,.
\label{equ_constr_invariance}
\vspace*{-2mm}
\end{equation}
\end{itemize}
We introduce these constraints (which are rather ad-hoc) because under them a closed-form solution of \eqref{equ_opt_problem_2}
can be obtained. This constrained solution is given
\vspace*{-1mm}
by \cite{SparseToAppear}
\begin{equation} \label{equ BBc notin supp}
\hat{x}_{\text{c},k}(\mathbf{y};\mathbf{x}_{0}) \eq y_{k} \Bigg[ 1 - \!\!\prod_{l \ist\in\ist \supp(\mathbf{x}_{0})} \!\!\!  \!\rmv \tanh\!\bigg( \frac{ y_l \ist x_{0,l}}{\sigma^{2}}\bigg) \Bigg]
\vspace*{-2mm}
\end{equation}
for $\mathbf{y} \!\in\! \mathbb{R}_{+}^{N}$. For other values of $\mathbf{y}$, the solution can be obtained via the properties \eqref{equ_constr_odd_symm}
and \eqref{equ_constr_invariance}.

Combining \eqref{equ BBc in supp} and \eqref{equ BBc notin supp}, we obtain a constrained solution of \eqref{equ_opt_problem_1} which will be denoted $\hat{\mathbf{x}}_{\text{c}}(\mathbf{y};\mathbf{x}_{0})$.
The MSE
of $\hat{\mathbf{x}}_{\text{c}}(\mathbf{y};\mathbf{x}_{0})$ at the parameter value $\mathbf{x}_{0}$ is the tightest lower bound
on the MSE
at $\mathbf{x}_{0}$ of any unbiased estimator that satisfies the constraints \eqref{equ_constr_odd_symm} and \eqref{equ_constr_invariance}.
It will thus be called the \emph{constrained BB} and denoted by $\mbox{BB}_{\text{c}}(\mathbf{x}_{0})$.
Note that $\mbox{BB}_{\text{c}}(\mathbf{x}_{0})$ is an upper bound on the BB.
A closed-form expression of $\mbox{BB}_{\text{c}}(\mathbf{x}_{0})$ is given in the next theorem.
\vspace*{-.3mm}
\begin{theorem}
\label{thm_upper}
The BB (i.e., the minimum MSE achievable by any unbiased estimator) at $\x_0$ is bounded above \vspace{-1.5mm}
by
\emph{
\begin{equation}
\label{equ_constr_BB}
 \mbox{BB}_{\text{c}}(\mathbf{x}_{0}) \eq S \sigma^{2} \rmv+ (N\!\rmv-\!S) \,\sigma^{2}
 \Bigg[ 1 - \!\!\prod_{l \ist\in\ist \supp(\mathbf{x}_{0})} \!\!\!  g(x_{0,l};\sigma^2) \Bigg] \,,    \nonumber
\vspace{-3.5mm}
\end{equation}
\emph{with}}
\vspace{-1mm}
\begin{align*}
&g(x;\sigma^2) \triangleq \nonumber\\
&\rule{5mm}{0mm}\frac{2}{\sqrt{2\pi\sigma^2}} \!\int_0^\infty \!\! e^{-(y^{2} +\ist x^{2})/(2\sigma^{2})}
  \sinh \!\bigg( \rmv \frac{y x}{\sigma^{2}} \rmv\bigg)
  \tanh \!\bigg( \rmv \frac{y x}{\sigma^{2}} \rmv\bigg)\ist dy \,.
\end{align*}
\vspace{-4mm}
\end{theorem}

We note that for all parameter values $\mathbf{x}_{0}$ with ${\| \mathbf{x}_{0} \|}_{0} \! <\! S$, the lower and upper bounds of Theorem \ref{thm_lower} and \ref{thm_upper} coincide and equal $N \sigma^{2}\rmv$, which is therefore also the BB. This value is identical to the unconstrained CRB. We refer to \cite{ZvikaSSP,ZvikaCRB} for a discussion of this result.

\vspace*{-1.8mm}

\section{Numerical Results} \label{sec_sim}

\vspace{-.9mm}

In Section~\ref{sec_optimum_unbiased_estimation}, we bounded
the achievable performance of unbiased estimators as a means for quantifying the difficulty of estimation in the SSNM\@. One use of this analysis is in the identification of the threshold region, a range of SNR values which constitutes a transition between low-noise and high-noise behavior. Specifically,
the performance of estimators can often be calculated analytically when the SNR $\|\x_0\|_2^2 / (N \sigma^2)$ is either very high or very low. It is then important to identify the threshold which separates these two regimes.

The lower and upper bounds on the BB which were derived above also exhibit a transition between a high-SNR region and a low-SNR region.
In the high-SNR region, the lower bound (assuming ${\|\x_0\|}_0 \rmv=\rmv S$) and the upper
bound both converge to $S \sigma^2\rmv$, while in the
low-SNR region,
both bounds are on the order of $N \sigma^2\rmv$. The true BB therefore also
displays such a threshold. Since the BB is itself a lower bound
\pagebreak 
on the MSE of unbiased estimators, one would expect that the transition region of the BB occurs at slightly lower SNR values than that of actual estimators.

To test this hypothesis, we compared the bounds of Section~\ref{sec_optimum_unbiased_estimation} with the MSE of two well-known \emph{biased}
estimation schemes. First, we considered the maximum likelihood (ML) estimator, which can be shown to
\vspace*{-1mm}
equal
\begin{equation}
\label{equ_ml_est}
\hat{x}_{\mathrm{ML},k}(\mathbf{y}) = \begin{cases}
 y_k ,  & \text{if } k \in \mathcal{L} \\
 0, & \text{else}
\end{cases}
\vspace*{-1mm}
\end{equation}
where $\mathcal{L}$ denotes the indices of the $S$ largest (in magnitude) entries of $\mathbf{y}$. We also considered the hard-thresholding (HT)
\vspace{-.2mm}
estimator
 \begin{equation}
 \label{equ_threshold_est}
 \hat{x}_{\mathrm{HT},k}(\mathbf{y})  =  \begin{cases}
 y_k ,  & \text{if } |y_k| \geq T \\
 0, & \text{else}
\end{cases}
\vspace*{-.5mm}
 \end{equation}
with the commonly employed threshold $T = \sigma \sqrt{2 \log{N}}$ \cite{Mallat98}.

\begin{figure}[t]
\vspace{-.9mm}
\centering
\psfrag{SNR}[c][c][1]{\uput{3.6mm}[270]{0}{SNR (dB)}}
\psfrag{MSE}[c][c][1.4][270]{\uput{0.1mm}[180]{0}{ $\frac{\varepsilon}{\sigma^{2}}$}}
\psfrag{x_0}[c][c][1]{\uput{0.3mm}[270]{0}{$0$}}
\psfrag{x_0_1}[c][c][.9]{\uput{0.3mm}[270]{0}{$-10$}}
\psfrag{x_0_01}[c][c][.9]{\uput{0.3mm}[270]{0}{$-20$}}
\psfrag{x_1}[c][c][.9]{\uput{0.3mm}[270]{0}{$0$}}
\psfrag{x_5}[c][c][1]{\uput{0.1mm}[270]{0}{$5$}}
\psfrag{x_8}[c][c][1]{\uput{0.1mm}[270]{0}{$8$}}
\psfrag{x_10}[c][c][.9]{\uput{0.3mm}[270]{0}{$10$}}
\psfrag{x_100}[c][c][.9]{\uput{0.3mm}[270]{0}{$20$}}
\psfrag{y_2}[c][c][1]{\uput{0.1mm}[180]{0}{}}
\psfrag{y_12}[c][c][.9]{\uput{0.1mm}[180]{0}{$12$}}
\psfrag{y_14}[c][c][1]{\uput{0.1mm}[180]{0}{}}
\psfrag{y_4}[c][c][.9]{\uput{0.1mm}[180]{0}{$4$}}
\psfrag{y_6}[c][c][.9]{\uput{0.1mm}[180]{0}{$6$}}
\psfrag{y_8}[c][c][.9]{\uput{0.1mm}[180]{0}{$8$}}
\psfrag{y_10}[c][c][.9]{\uput{0.1mm}[180]{0}{$10$}}
\psfrag{ML}[l][l][0.8]{$\varepsilon_{\mathrm{ML}}$}
\psfrag{HCRB}[l][l][0.8]{HCRB}
\psfrag{BB}[l][l][0.8]{$\mbox{BB}_{\text{c}}$}
\psfrag{Hard}[l][l][0.8]{$\varepsilon_{\mathrm{HT}}$}
\psfrag{CRB}[l][l][0.8]{CRB}
\centering
\hspace*{3mm} \includegraphics[height=3.77cm,width=8.2cm]{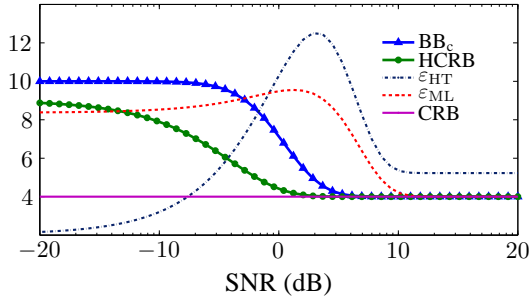}
\vspace{-2mm}
  \caption{MSE of the HT and ML estimators compared with the performance bounds $\mbox{BB}_{\text{c}}$, HCRB and CRB
  (all normalized by $\sigma^{2}$), as a function of the SNR.}
\label{fig_all_equal}
\vspace{-.5mm}
\end{figure}

We used $N \!=\! 10$ and $S \!= 4$, and chose $50$ parameter vectors $\x_0$ from the set $\mathcal{R} \triangleq \{ c \, (1,1,1,1,0,0,0,0,0,0)^{T} \}_{c \in \mathbb{R}_{+}}$,
with different values of $c$ to obtain a wide range of SNR values. The results are plotted in Fig.~\ref{fig_all_equal}.
Although there is some gap between the lower bound (HCRB) and the upper bound ($\mbox{BB}_{\text{c}}$), a rough indication of the behavior of the BB is conveyed.
As expected, the SNR threshold predicted by these bounds is somewhat lower than that of practical estimators. Specifically, the transition region of the BB can be seen to occur at SNR values between $-5$ and $5$ dB, while the transition of the ML and HT estimators is at SNR values between $0$ and $10$ dB\@.
Another effect which is
visible in Fig.~\ref{fig_all_equal} is the convergence of the ML estimator to the BB at high SNR; this is a consequence of the well-known fact that the ML approach is asymptotically unbiased and asymptotically minimizes the MSE at high SNR. Furthermore, our performance bounds in Fig.~\ref{fig_all_equal} suggest that at intermediate and high SNR (above $0$ dB),
there may exist unbiased estimators that outperform the ML and HT estimators.
However, at low SNR, both estimators (ML and HT) are better than the best unbiased estimator. This agrees with the general rule that unbiased estimators perform poorly at low SNR.
\begin{figure}[t]
\vspace{-1mm}
\centering
\psfrag{SNR}[c][c][1]{\uput{3.0mm}[270]{0}{SNR (dB)}}
\psfrag{Tightness}[c][c][1][270]{\uput{0.8mm}[180]{0}{}}
\psfrag{x_0_01}[c][c][.9]{\uput{0mm}[270]{0}{$-20$}}
\psfrag{x_0_1}[c][c][.9]{\uput{0mm}[270]{0}{$-10$}}
\psfrag{x_1}[c][c][.9]{\uput{0mm}[270]{0}{$0$}}
\psfrag{x_10}[c][c][.9]{\uput{0mm}[270]{0}{$10$}}
\psfrag{x_100}[c][c][.9]{\uput{0mm}[270]{0}{$20$}}
\psfrag{x_1000}[c][c][.9]{\uput{0mm}[270]{0}{$30$}}
\psfrag{y_2}[c][c][1]{\uput{0.1mm}[180]{0}{}}
\psfrag{y_1}[c][c][.9]{\uput{0.1mm}[180]{0}{$1$}}
\psfrag{y_1_2}[c][c][.9]{\uput{0.1mm}[180]{0}{$1.2$}}
\psfrag{y_1_4}[c][c][.9]{\uput{0.1mm}[180]{0}{$1.4$}}
\psfrag{y_1_6}[c][c][.9]{\uput{0.1mm}[180]{0}{$1.6$}}
\psfrag{y_1_8}[c][c][.9]{\uput{0.1mm}[180]{0}{$1.8$}}
\psfrag{data2}[l][l][0.7]{$\mathcal{R}_{2}$}
\psfrag{data1}[l][l][0.7]{$\mathcal{R}$}
\psfrag{data3}[l][l][0.7]{$\mathcal{R}_{3}$}
\centering
\hspace*{1mm} \includegraphics[height=3.78cm,width=8.3cm]{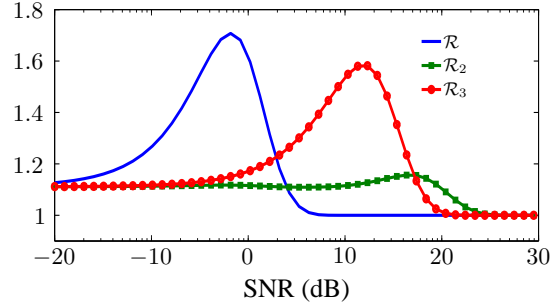}
\vspace{1.9mm}
  \caption{Ratio $\mbox{BB}_{\text{c}}(\mathbf{x}_{0}) / \mbox{HCRB}(\mathbf{x}_{0})$ for different sets of parameter vectors $\mathbf{x}_{0}$.}
\label{fig_tightness}
\vspace{-.5mm}
\end{figure}

One may argue that considering only parameter values in the set $\mathcal{R}$ is
not representative, since $\mathcal{R}$
covers only a small part of the parameter space $\mathcal{X}_S$.
\pagebreak 
However, it can be shown that
the choice for $\mathcal{R}$ is conservative in that the maximum deviation between the HCRB and the $\mbox{BB}_{\text{c}}$ is largest
when the non-zero entries of $\mathbf{x}_{0}$ have approximately the same magnitude, which is the case for each element of $\mathcal{R}$.
To illustrate this fact, we considered the two additional sets $\mathcal{R}_{2} \triangleq \{ c \, (0.1,1,1,1,0,0,0,0,0,0)^{T} \}_{c \in \mathbb{R}_{+}}$ and $\mathcal{R}_{3} \triangleq \{ c \, (10,1,1,1,0,0,0,0,0,0)^{T} \}_{c \in \mathbb{R}_{+}}$,
in which the four non-zeros are not all equal. Fig.~\ref{fig_tightness} depicts the ratio $\mbox{BB}_{\text{c}} / \mbox{HCRB}$
versus the SNR for the three sets $\mathcal{R}$, $\mathcal{R}_2$, and $\mathcal{R}_3$. It is seen that the maximum value of $\mbox{BB}_{\text{c}} / \mbox{HCRB}$
is indeed highest when $\x_0$ is in $\mathcal{R}$.

\vspace{-2.5mm}
\section{Conclusion}
\vspace{-1.4mm}
We considered unbiased estimation of a sparse parameter vector in white Gaussian noise.
The Barankin bound (i.e., the minimum MSE achievable by any unbiased estimator) was
characterized via upper and lower bounds.
Our numerical results suggest that these bounds may also give information about the success of more general, biased techniques,
for example by providing a lower bound on the threshold region.
A subject for further study is the extension of these results to a model of the form $\mathbf{y} = \mathbf{A} \mathbf{x}_0 + \mathbf{n}$ in which $\x_0$ is sparse and $\mathbf{A}$ is an arbitrary
matrix with full column rank.

\end{document}